\documentclass[showpacs,amsmath,amssymb,twocolumn,prl,notitlepage]{revtex4-1}

\usepackage{qcircuit}
\usepackage[dvips]{graphicx}
\usepackage{amsmath,amssymb,amsthm,mathrsfs,amsfonts,dsfont}
\usepackage{subfigure, epsfig}
\usepackage{braket}
\usepackage{bm}
\usepackage{enumerate}
\usepackage{color}
\usepackage{multirow}
\usepackage{fancybox, graphicx} 
\usepackage[skins]{tcolorbox}

\newcommand{\tr}{\mathrm{Tr}}
\newcommand{\veps}{\varepsilon}

\newcommand{\N}{\mathcal{N}}
\newcommand{\U}{\mathcal{U}}
\newcommand{\V}{\mathcal{V}}
\newcommand{\C}{\mathcal{C}}
\newcommand{\D}{\mathcal{D}}
\newcommand{\I}{\mathcal{I}}
\newcommand{\E}{\mathcal{E}}
\newcommand{\M}{\mathcal{M}}
\newcommand{\Se}[2]{S\left({#1}\|{#2}\right)}

\newtcolorbox[auto counter]{tbox}[2][]{%
    enhanced, float=hbt, drop fuzzy shadow southeast,
    colback=white!5!white, colframe=white!50!black,
    width= .99\columnwidth,sharp corners,boxrule=0.8pt,
    title={Properties of the quantum relative entropy of channels. #2}, #1
}

\begin{document}

\title{Relative entropies of quantum channels with applications in resource theory}
\begin{abstract}
Entropic quantifiers of states lie at the cornerstone of the quantum information theory. 
While a quantum state can be abstracted as a device that only has outputs, the most general quantum device is a quantum channel that also has inputs.  
In this work, we extend the entropic quantifiers of states to the ones of channels.
In the one-shot and asymptotic scenarios, we propose relative entropies of channels under the task of hypothesis testing. 
Then, we define the entropy of channels based on relative entropies from the target channel to the completely depolarising channel.  
We also study properties of relative entropies of channels and the interplay with entanglement. 
Finally, based on relative entropies of channels, we propose general resource theories of channels and discuss the coherence of general channels and measurements, and the entanglement of channels.
\end{abstract}

\date{\today}

\author{Xiao Yuan}
\email{xiao.yuan.ph@gmail.com}
\affiliation{Department of Materials, University of Oxford, Parks Road, Oxford OX1 3PH, United Kingdom}

\maketitle

The quantum information theory is based on definitions of entropic quantifiers \cite{nielsen2010quantum}, which enable to construct resource theories \cite{chitambar2018quantum} and quantify the performance of quantum protocols, such as quantum key distribution \cite{devetak2005distillation,coles2016numerical}, quantum random number generation \cite{YuanPhysRevA2015,yuan2016interplay,PhysRevA.97.012302}, and quantum computing \cite{veitch2014resource,PhysRevX.6.021043, howard2014contextuality}. Especially, the quantum relative entropy of two states \cite{umegaki1962conditional} is defined by $S(\rho\|\sigma) = \tr[\rho\log(\rho)-\rho\log(\sigma)]$, which measures the power of quantum hypothesis testing \cite{RevModPhys.74.197}. The quantum relative entropy is an important measure for describing the difference between two quantum states and is a fundamental tool that has been used in studying entropic properties of quantum states \cite{nielsen2010quantum,RevModPhys.74.197}, quantifying coherence \cite{Baumgratz14,RevModPhys.89.041003} and entanglement \cite{PhysRevLett.78.2275,RevModPhys.81.865}, investigating quantum thermodynamics \cite{lostaglio2015description}, etc. 
Especially, considering the quantum relative entropy between any state $\rho$ and the maximally mixed state $I_A/d_A$, we can also define the Von-Neumann entropy of $\rho$ by $S(\rho) = \log_2d_A - S(\rho\|I_A/d_A)$, which can be further extended to define other entropic measures.

Although quantum states describe the status of physics systems, the most general physics objects are quantum channels that map input systems to output systems. A quantum channel reduces to a quantum state or a demolition measurement when the input system is null or the output system is purely classical, respectively. Many properties of quantum channels have been extensively studied, such as the Holevo information \cite{holevo1998capacity}, the mutual information of a channel \cite{PhysRevA.56.3470}, and the general quantum channel capacity \cite{2018arXiv180102019G}. Distance measures of two channels have been proposed  \cite{PhysRevA.71.062310} and applied in channel discrimination \cite{PhysRevLett.102.250501}, studying the non-Markovianity of quantum processes \cite{PhysRevA.81.062115,rivas2014quantum}, etc. However, most works still focus on the properties of quantum states and regard quantum channels as processes of states. This leads to the fact that many important properties, such as entropic quantifiers, of quantum channels are not well studied. 

Meanwhile, resource theories are to investigate the characterisation, quantification, and manipulation of resources \cite{chitambar2018quantum}. 
Many works have been focused on the resource theory of states such as coherence \cite{Baumgratz14,RevModPhys.89.041003}, entanglement \cite{PhysRevLett.78.2275,RevModPhys.81.865}, and thermodynamics \cite{PhysRevLett.111.250404,horodecki2013fundamental,Brandao3275}, and the state induced resources such as nonlocality \cite{RevModPhys.86.419} and contextuality \cite{PhysRevA.71.052108,liang2011specker}. 
Recently, the coherence resource theory is extended to quantum channels to characterise the coherence of operations \cite{theurer2018quantifying} by the trace distance measure of operations \cite{PhysRevA.71.062310}.  
However, the basic manipulation processes, including distillation and dilution, are general quantified by entropic quantifiers.  Because entropic quantifiers of channels are not well studied, resource theories of channels are yet to be investigated.

In this work, we first propose entropic quantifiers of quantum channels. Specifically, we investigate the relative entropy of channels that is firstly discussed in Refs.~\cite{cooney2016strong,PhysRevA.97.012332}.
Focusing on hypothesis testing, we extend the previous definition to other scenarios by considering different classes of input states.
We also define entropies of channels via relative entropies from the channel to the completely depolarising channel, which is also independently proposed by Gour and Wilde recently \cite{gour2018entropy}.  
Then, we investigate the role of entanglement in hypothesis testing of quantum channels and study the properties that the definitions should satisfy. 
With relative entropies of channels, we propose general resource theories of channels and study the coherence of general channels and measurements, and the entanglement of channels.

\emph{Relative entropy of two quantum channels.---}
Quantum relative entropy of states measures the hypothesis testing power.
Suppose the initial hypothesis is $\sigma$, but the actual state is $\rho$. When a measurement is performed on a single copy of state $\rho$, with failure probability less than $\varepsilon$, the probability or the $p$-value that the initial hypothesis is true  is lower bounded by $
	p = 2^{-D_{H}^\varepsilon(\rho\|\sigma)}$ \cite{Wang12}.
Here, the one-shot hypothesis testing relative entropy \cite{buscemi2009quantum} is defined by
$D_H^\varepsilon(\rho\|\sigma) = -\log_2 \min_{Q:0\le Q\le I, \tr[Q\rho]\ge1-\veps}\tr[Q\sigma]$,
with $\{Q,I-Q\}$ being the measurement that distinguishes between $\rho$ and $\sigma$.
When measuring $\rho$, the probabilities of obtaining $Q$ and $I-Q$ are $1-\varepsilon$ and $\varepsilon$, respectively. Therefore, with failure probability $\varepsilon$, we can disprove the hypothesis $\sigma$ when the outcome $I-Q$ is obtained by measuring $\sigma$, and the probability that the hypothesis is true is $\tr[Q\sigma]$.

\begin{figure}[b]\centering
{\includegraphics[width=8.5cm]{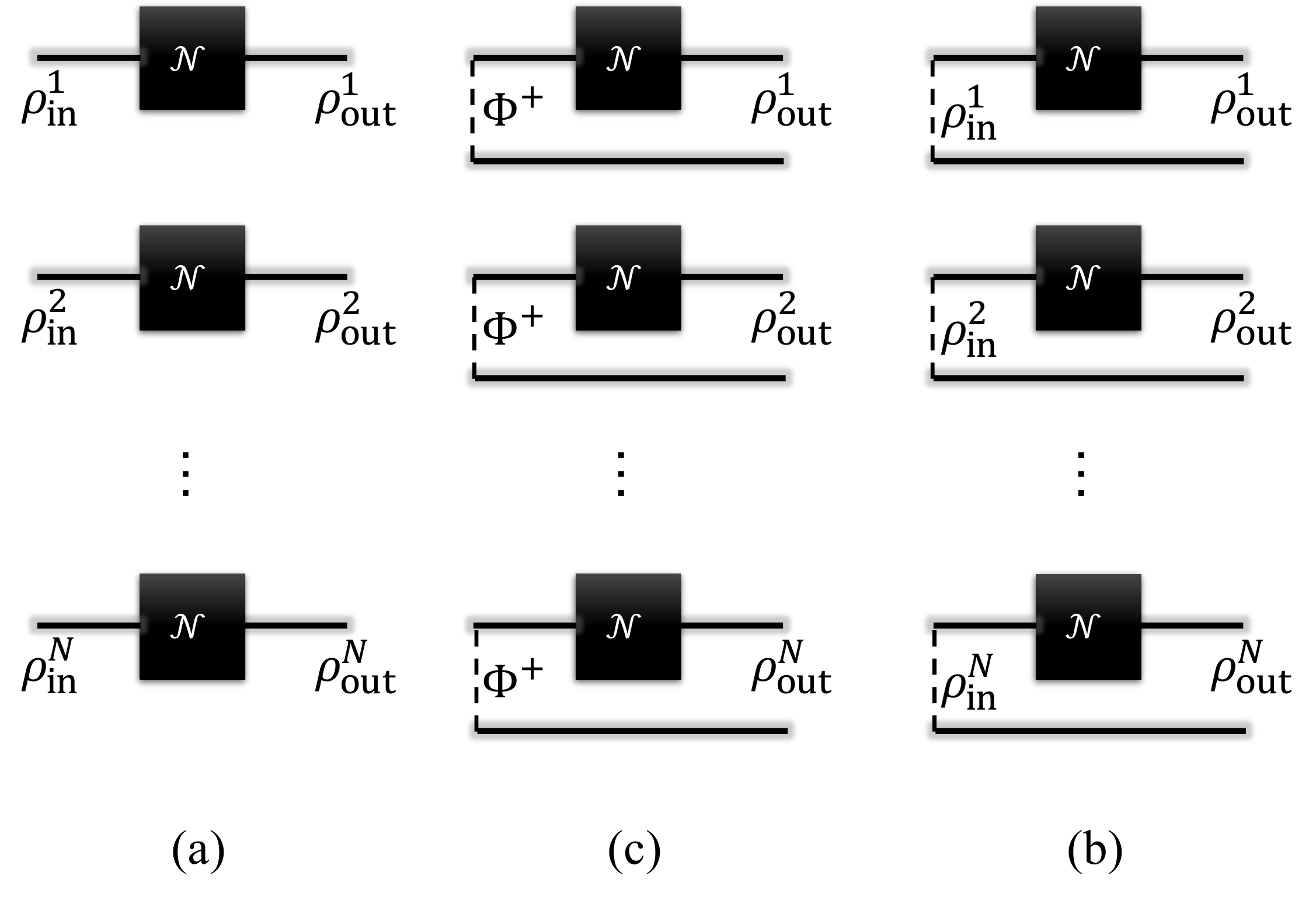}}
 \caption{Hypothesis testing of two channels. (a) For each use of the quantum channel, no additional ancilla is allowed. (b) One party of the maximally entangled state is input to the channel. (c) One party of a joint state is input to the channel.  } \label{Fig:hypothesis0}
\end{figure}

Now, we introduce the hypothesis testing of channels.
Focus on quantum channels $\N_{A'\rightarrow A}$ and $\M_{A'\rightarrow A}$ that map system $A'$ with dimension $d_{A'}$ to system $A$ with dimension $d_{A}$. For simplicity, we omit the subscript in the following when there is no confusion. Suppose the initial hypothesis is that the quantum channel is $\M$, albeit it is actually described by $\N$. Given $N$ uses of the channel $\N$, we test the correctness of the initial hypothesis. 
We consider three different types of identical and independent inputs as shown in Fig.~\ref{Fig:hypothesis0}: (a) the input state is only system $A'$; (b) the input state is the maximally entangled state $\ket{\Phi^+} = 1/\sqrt{d_{A'}}\sum_i\ket{ii}$ of system $A'$ and another ancillary system $B$; (c) the input state is a general entangled state $\psi_{A'B}$ of systems $A'$ and any ancillary system $B$. For each types of input states, we minimise the $p$-value of hypothesis $\M$ over all possible input states.

For a single use of the channel, we define the relative entropies of channels $\N$ and $M$ as follows.

\noindent\emph{Definition 1. The $A$-, $\Phi^+$-, and $AB$-one-shot relative entropies of two channels $\N$ and $\M$ are respectively, }
	\begin{equation}
\begin{aligned}\nonumber
	D^\varepsilon_{A}({\N}\|{\M}) &= \max_\psi D_H^\varepsilon(\N(\psi)\|\M(\psi)),\\
	D^\varepsilon_{\Phi^+}(\N\|\M) &=D_H^\varepsilon((\N\otimes\I)(\Phi^+)\|(\M\otimes\I)(\Phi^+)),\\
		D^\varepsilon_{AB}({\N}\|{\M}) &= \max_{\psi_{A'B}} D_H^\varepsilon((\N\otimes\I)(\psi_{A'B})\|(\M\otimes\I).
\end{aligned}
	\end{equation}
Note that $(\N\otimes\I)(\Phi^+)$ is the Choi state of channel $\N$, the relative entropy $D^\varepsilon_{\Phi^+}({\N}\|{\M})$ is the relative entropy of Choi states.
It is also straightforward to see that 
$D^\varepsilon_{AB}({\N}\|{\M}) \ge \max\{D^\varepsilon_{A}({\N}\|{\M}),D^\varepsilon_{\Phi^+}({\N}\|{\M})\}$.
While we will show later that the equal sign cannot be achieved for all channels and there is no such a definite order between $D^\varepsilon_{A}({\N}\|{\M})$ and $D^\varepsilon_{\Phi^+}({\N}\|{\M})$. In this work, we focus on the measures with $\varepsilon=0$ and omit $\varepsilon$ afterwards.

Next, we consider the asymptotic case where the channel is used $N\rightarrow\infty$ times. According to the quantum Stein's lemma \cite{Hiai1991,Ogawa00},
$\lim_{N\rightarrow \infty}\frac{1}{N}D_{H}^\veps(\rho^N\|\sigma^N) = S(\rho\| \sigma)$, we can alternatively define relative entropies of channels as follows.

\noindent\emph{Definition 2. The $A$-, $\Phi^+$-, and $AB$-relative entropies of two channels $\N$ and $\M$ are respectively,}
	\begin{equation}
\begin{aligned}\nonumber
S_{A}({\N}\|{\M}) &= \max_\psi S(\N(\psi)\|\M(\psi)),\\
S_{\Phi^+}(\N\|\M) &=S((\N\otimes\I)(\Phi^+)\|(\M\otimes\I)(\Phi^+)),\\
S_{AB}({\N}\|{\M}) &= \max_{\psi_{A'B}} S((\N\otimes\I)(\psi_{A'B})\|(\M\otimes\I)(\psi_{A'B})).
\end{aligned}
\end{equation}
Similarly, $S_{AB}({\N}\|{\M}) \ge \max\{S_{A}({\N}\|{\M}),S_{\Phi^+}({\N}\|{\M})\}$, and there is no definite order between $S_{A}({\N}\|{\M})$ and $S({\N}\|{\M})$.
The relative entropy of $S_{AB}({\N}\|{\M})$ was firstly proposed in Ref.~\cite{cooney2016strong} and the generalisations with generalised divergence were studied in Ref.~\cite{PhysRevA.97.012332}.

Consider special cases where $\M$ is the completely depolarising channel, $\D(\sigma) = I_A/d_{A}$, with $I_A$ being the identity matrix of system $A$.
For the one-shot relative entropy $D_H(\rho\|\sigma)$ and the relative entropy $S(\rho\|\sigma)$, we have $D_H(\rho\|I_A/d_A) = \log_2d_A - S_0(\rho)$ and $S(\rho\|I_A/d_A) = \log_2d_A-S(\rho)$, respectively. Here $S_0(\rho)$ and $S(\rho)$ are the R\'enyi entropy $S_\alpha=\frac{1}{1-\alpha}\log_2\tr[\rho^\alpha]$ with $\alpha\rightarrow0$ and $\alpha\rightarrow1$ and, respectively.
The relative entropies between the channel $\N$ and the completely depolarising channel $\D$ are 
\begin{equation}\label{Eq:ioneshot}
\begin{aligned}\nonumber
D_A(\N\|\D) &= \log_2d_A - \min_{\psi}S_0(\N(\psi)),\\
D_{\Phi^+}(\N\|\D) &= 2\log_2d_A - S_0(\N\otimes\I(\Phi^+)),\\
D_{AB}(\N\|\D) &=\max_{\psi_{A'B}} D_H^\varepsilon((\N\otimes \I) (\psi_{A'B})\|\frac{I_A}{d_A}\otimes \tr[\psi_{A'B}]),\\
S_A(\N\|\D) &= \log_2d_A - \min_{\psi}S(\N(\psi)),\\
S_{\Phi^+}(\N\|\D) &= 2\log_2d_A - S(\N\otimes\I(\Phi^+)),\\
S_{AB}(\N\|\D) &= \log_2d_A - \min_{\psi_{A'B}}H(A|B)_{\rho_{AB} =\N\otimes\I(\psi_{A'B})}.\\
\end{aligned}
\end{equation}
And we can define the entropy of quantum channels. 

\noindent\emph{Definition 3. 	Given the relative entropy $S(\N\|\M)$, the entropy of a quantum channel $\N$ is }
\begin{equation}
	S(\N) = \log_2d_A - S(\N\|\D).
\end{equation}
Here, $S(\N\|\M)$ denotes one of the six relative entropy definitions.
Especially, for $S_{A}(\N\|\D)$ and $S_{AB}(\N\|\D)$, we have $S_{A}(\N) = \min_{\psi}S(\N(\psi))$ and $S_{AB}(\N) = \min_{\psi_{A'B}}H(A|B)_{\rho_{AB} =\N\otimes\I(\psi_{A'B})}$, respectively. 
When $A'$ is a null system, $\N$ is a state preparation channel, $\N(\emptyset) = \rho$, and the entropies $S_{A}(\N)$ and $S_{AB}(\N)$ reduce to the entropy of state $\rho$, $S_{A}(\N)=S_{AB}(\N) = S(\rho)$.
Recently, the entropy of channels $S_{AB}(\N)$ is independently proposed by Gour and Wilde \cite{gour2018entropy}, who also studied its properties and its operational meaning in quantum channel merging. 

\emph{Entanglement in hypothesis testing.---}
The three different scenarios in Fig.~\ref{Fig:hypothesis0} correspond to three different cases in hypothesis testing where the input is not entangled, maximally entangled, and generally entangled, respectively.
Now, we study the role of entanglement in hypothesis testing by comparing the relative entropies.
Specifically, we consider whether relative entropies $S(\N\|\M)$ of channels with general entangled states is strictly larger than the ones with maximally entanglement and the ones without entanglement. 
We first consider the case that $\M$ is the completely depolarising channel $\D$.
When $\N$ is the identity channel $\I(\rho)=\rho$, we can show that $D_{AB}({\I}\|{\D})=D_{\Phi^+}({\I}\|{\D}) = S_{AB}({\I}\|{\D}) = S_{\Phi^+}(\I\|\D) = 2\log_2d_A$ and $D_{A}(\I\|\D) =S_{A}(\I\|\D) =\log_2d_A$. We have $D_{\Phi^+}(\I\|\D)>D_{A}(\I\|\D)$ and $S_{\Phi^+}(\I\|\D)>S_{A}(\I\|\D)$.
Therefore, inputting entangled state is strictly stronger than inputting without the entangled ancilla.

Consider a qubit channel $\N_{0}$ with Kraus operators $K_0 = \sqrt{0.5}\ket{0}\bra{0}$, $K_1 = \sqrt{0.5}\ket{1}\bra{0}$, and $K_2 = \ket{1}\bra{1}$. Then $D_{AB}(\N_{0}\|\D)=S_{AB}(\N_{0}\|\D) = D_{A}(\N_{0}\|\D) =S_{A}(\N_{0}\|\D) = 1$ with input state $\ket{\psi} = \ket{1}$. While when the input state is $\ket{\Phi^+_2}=1/\sqrt{2}(\ket{00}+\ket{11})$, we have $D_{\Phi^+}({\N_{0}}\|{\D}) = \log_2(4/3)$ and $S_{\Phi^+}({\N_{0}}\|{\D}) = 1/2$. 
Therefore, for channel $\N_{0}$, we have $D_{\Phi^+}(\N_{0}\|\D)<D_{A}(\N_{0}\|\D)=D_{AB}(\N_{0}\|\D)$ and $S_{\Phi^+}(\N_{0}\|\D)<S_{A}(\N_{0}\|\D)=S_{AB}(\N_{0}\|\D)$.  With the two examples, we conclude that the strategy of Fig.~\ref{Fig:hypothesis0}(c) are generally stronger than the strategies of Fig.~\ref{Fig:hypothesis0}(a) and Fig.~\ref{Fig:hypothesis0}(b); while the strategies between Fig.~\ref{Fig:hypothesis0}(a) and Fig.~\ref{Fig:hypothesis0}(b) are not superior to each other. Therefore, there exist non-maximally entangled states that can maximise the hypothesis testing probability of two channels.

\emph{Properties.---}
Now, we consider the properties that the quantum relative entropy $\Se{\N}{\M}$ should satisfy.
Firstly, because the relative entropy measures the difference between two objects, it should be non-negative. Other properties arise from manipulating a single channel or a set of channels. 
Consider a single channel transform by independently applying additional channels before and after the channel, 
weak monotonicity requires the relative entropy to be non-increasing under such a transform. Especially, when applying reversible operations $\U_1$ and $\U_2$ before and after the channels, respectively, the relative entropy is invariant
i.e., $S(\mathcal{V}_1\circ\N\circ\mathcal{V}_2\|\mathcal{V}_1\circ\M\circ\mathcal{V}_2)= S(\N\|\M)$.  A superchannel \cite{chiribella2008transforming} transforms a channel $\N_{A'\rightarrow A}$ to $\N'_{C'\rightarrow C}=\Phi(\N_{A'\rightarrow A}) = \V'_{AE\rightarrow C}\circ(\N_{A'\rightarrow A}\otimes\I_E)\circ\V_{C'\rightarrow A'E}$, with ancillary system $E$, and channels  $\V'_{AE\rightarrow C}$ and $\V_{AE\rightarrow C}$.
The strong monotonicity requires the relative entropy to be non-increasing under any superchannel.

With a set of quantum channels, a new channel can be obtained by probabilistically applying the channels. It is called jointly convex when the relative entropy cannot be increased by such mixing operations.
Quantum channels can be also applied jointly. As the input and output systems are the tensor product of the input and output of the two channels, the relative entropy should not decrease by adding two channels according to the additivity property. 
We call it strictly additive when the relative entropy is the same after adding channels. 
In the additivity requirement, we can set $\N_0$ and $\M_0$ to be the identity channel. The relative entropy is called stable when it is invariant by combining with the identity channel.
\begin{enumerate}[S1]
	\item (\emph{Non-negativity}) The relative entropy is non-negative, $S(\N\|\M)\ge 0$. The equality sign hods iff $\N\equiv\M$.
	\item (\emph{Weak monotonicity}) The relative entropy is non-increasing by sandwiching it with other channels $
		S(\mathcal{V}_1\circ\N\circ\mathcal{V}_2\|\mathcal{V}_1\circ\M\circ\mathcal{V}_2)\le S(\N\|\M)$. \\
		(\emph{Strong monotonicity}) The relative entropy is non-increasing under superchannels $\Phi$, i.e.,
	$S(\Phi(\N)\|\Phi(\M))\le S(\N\|\M)$.
	\item (\emph{Joint convexity}) The relative entropy is jointly convex $S\left(\sum_ip_i\N_i\big\|\sum_ip_i\M_i\right)\le \sum_i p_iS(\N_i\|\M_i)$. 
	\item (\emph{Additivity}) The additivity property requires
 	$\Se{\N_0 \otimes \N_1}{\M_0 \otimes \M_1}	 \ge  \Se{\N_0}{\M_0} + \Se{\N_1}{\M_1}$.
	\item (\emph{Stability}) The stability property requires
 	$\Se{\I \otimes \N}{\I \otimes \M}	 =\Se{\N}{\M}$.
\end{enumerate}

In Table \ref{Table:properties}, we summarise properties of the six relative entropy definitions. 
The relative entropies based on Choi matrices of channels are easy to calculate, but violate both the strong and weak monotonicity properties. The $A$-quantum relative entropies also violate the strong monotonicity and stability. The $AB$-relative entropies satisfy all the properties, albeit they are hard to calculate due to the maximisation in the definition.

\begin{table*}[t]
\centering
  \caption{Properties of the proposed quantum relative entropy measures for quantum channels.}
\begin{tabular}{ccccccccc}
\hline
\multirow{2}{*}{Measures}&\multirow{2}{*}{Non-negativity}&\multicolumn{2}{c}{Monotonicity}&\multirow{2}{*}{Joint convexity}&\multirow{2}{*}{Additivity}&\multirow{2}{*}{Stability}\\
&&Weak&Strong&&&&
\\
\hline
$D_A(\N\|\M)$, $S_{AB}(\N\|\M)$&Yes&Yes&No&Yes&Yes, not strict&No\\
$D_{\Phi^+}(\N\|\M)$, $S_{\Phi^+}(\N\|\M)$&Yes&No&No&Yes&Yes, strict&Yes\\
$D_{AB}(\N\|\M)$, $S_{AB}(\N\|\M)$&Yes&Yes&Yes&Yes&Yes&Yes\\
\hline
\end{tabular}
\label{Table:properties}
\end{table*}

\emph{Quantum channel resource theory.---}
With the definitions of relative entropies of channels and the properties, we study general resource theories of channels, which are defined by resource free channels, resource free operations, and resource measures. 
Denote a resource free channel as $\C$ and the set of $\C$ as $S_{\C}=\{\C\}$. The maximal set $S_{\Phi}$ of resource free operations consists of superchannels $\Phi$ that map a resource free channel to another one, $\Phi(\C) = \C'\in S$. Other requirements of $\Phi$ may be applied for specific resource theories. Resource measures are  real-valued functions $Q$ of channels $\N$, which generally satisfy two properties. (P1) It is non-negative for general channels and vanishes for resource free channels, $Q(\N)\ge 0$ and $Q(\C)=0,\forall \C\in S_\C$; (P2) It cannot be increased under resource free operations, $Q(\N)\ge Q(\Phi(\N))$, $\forall \Phi\in S_{\Phi}$. Some other properties may be also required. For example, when $S_{\C}$ is convex, we can require that (P3) $Q(\N)$ cannot be increased under mixing, $\sum_ip_iQ(\Phi(\N_i))\ge Q(\sum_ip_i\N_i)$, for channels $\{\N_i\}$ and normalised probability distribution $\{p_i\}$. 

For a general channel resource theory, we define resource measures via relative entropies,
\begin{equation}
	Q_{rel}(\N) = \min_{\C\in S_\C} S(\N\|\C).
\end{equation}
According to the basic properties of relative entropies of channels, we show that $Q_{rel}(\N)$ satisfy (P1) and (P3) for all the six relative entropy definitions. However, because only $D_{AB}$ and $S_{AB}$ satisfy the strong monotonicity, only the definitions based on $D_{AB}$ or $S_{AB}$ satisfy the monotonicity requirement (P2). In the following, we discuss the coherence of general channels and measurements, and entanglement of channels with $S_{AB}$ as examples. The results follows similar for $D_{AB}$ and all the proofs can be found in Supplementary Materials. 

The coherence resource framework of channels was recently proposed \cite{theurer2018quantifying} based on the trace distance measure of channels \cite{PhysRevA.71.062310}.
Consider a channel $\N$ that maps system $A'$ to system $A$ with computational bases $I_{A'} = \{\ket{{i}_A}\}$ and $I_A = \{\ket{{i}_A}\}$, respectively.
Denote $\Delta_{A'}(\rho_{A'}) = \sum_{i_{A'}}\bra{i_{A'}}\braket{i_{A'}|\rho_{A'}|i_{A'}}\ket{i_{A'}}$ and $\Delta_A(\rho_A) = \sum_{i_A}\bra{i_A}\braket{i_A|\rho_A|i_A}\ket{i_A}$ to be the completely dephasing channels on system $A'$ and $A$, respectively. Three different classes of resource free states are defined \cite{theurer2018quantifying} based on resource destroying maps \cite{PhysRevLett.118.060502}, $\Delta_{A'}(\rho_{A'})$ and $\Delta_{A}(\rho_{A})$. 
A channel $\C_d$ is called detection-incoherent when $\Delta_A\circ\C_d=\Delta_A \circ\C_d\circ \Delta_{A'}$. 
A channel $\C_c$ is called creation-incoherent when $\C_c\circ\Delta_{A'}=\Delta_A \circ\C_c\circ \Delta_{A'}$. 
A channel $\C_{dc}$ is called detection-creation-incoherent when $\Delta_A\circ\C_{dc}=\C_{dc}\circ \Delta_{A'}$. 
For each class of resource free states, it corresponds to one resource theory. 
Suppose resource free operations are operations that map resource free channels to resource free channels. Then, the channel relative entropy of coherence can be defined by
\begin{equation}
	C_{rel}^{x}(\N) = \min_{\C_x} S_{AB}(\N\|\C_x),
\end{equation}
where $x$ denotes $d$, $c$, and $dc$.
Our definition is slightly different from the resource framework in Ref.~\cite{theurer2018quantifying}, which defines resource free operations by a sequential and/or parallel concatenation with resource free channels and allows system $A'$ and $A$ to change sizes. 
We compare the two definitions and prove the relative entropy measures for both definitions in Supplementary Materials. 

Next, we consider a special subset of channels, measurements. In analogy, a measurement $\M_d$ is called detection-incoherent when $\M_d(\rho) = \sum_k\tr[M_k\rho]\ket{\psi_k}\bra{\psi_k}$, with $M_k = \sum_{i_{A'}} p_{i_{A'}}^k\ket{i_{A'}}\bra{i_{A'}}$. 
A measurement $\M_c$ is called creation-incoherent when $\M_c(\rho) = \sum_{i_A}\tr[M_{i_A}\rho]\ket{{i_A}}\bra{{i_A}}$.
A measurement $\M_c$ is called detection-creation-incoherent when $\M_{dc}(\rho) = \sum_{i_A}\tr[M_{i_A}\rho]\ket{{i_A}}\bra{{i_A}}$ with $M_{i_A} = \sum_{i_{A'}} p_{i_{A},i_{A'}}\ket{i_{A'}}\bra{i_{A'}}$. Here, the POVM elements also satisfy $M_k\ge 0$, and $\sum_kM_k=I_{A'}$. Suppose resource free operations are operations that map resource free channels to resource free channels, the measurement relative entropy of coherence can be defined by
\begin{equation}
	C_{rel}^{x}(\N) = \min_{\M_x} S_{AB}(\N\|\M_x),
\end{equation}
where $x$ denotes $d$, $c$, and $dc$. Consider qubit projective measurement, $
	\N(\rho) = \braket{\psi_0|\rho|\psi_0}\ket{\psi_0}\bra{\psi_0}+\braket{\psi_1|\rho|\psi_1}\ket{\psi_1}\bra{\psi_1}$,
with normalised basis $\{\ket{\psi_0},\ket{\psi_1}\}$. We explicitly calculate the measurement relative entropies of coherence 
\begin{equation}
\begin{aligned}
	 C_{rel}^d(\N) &\in [C_{\min}(\ket{\psi_0}),C_{rel}(\ket{\psi_0})], \\
	C_{rel}^c(\N) &= C_{rel}^{dc}(\N) = C_{rel}(\ket{\psi_0}),
\end{aligned}
\end{equation}
where $C_{\min}(\ket{\psi_0})$ and $C_{rel}(\ket{\psi_0})$ are the min-entropy and the relative entropy of coherence of state $\ket{\psi_0}$,
$C_{\min}(\ket{\psi_0}) = H_{\min}(\psi_0^{diag})$ and 
		$C_{rel}(\ket{\psi_0}) = S(\psi_0^{diag})$, respectively.
Here $H_{\min}$ and $S$ are the min entropy and relative entropy of states, and $\psi_0^{diag}=\Delta_{A'}({\psi_0})$. Therefore, the coherence of qubit measurement channels is related to the coherence of the measurement basis states. 

Apart from coherence, we can also extend the entanglement theory to channels by focusing on channels $\N_{AC}$ that map systems $A'C'$ to systems $AC$. Note that the entanglement  theory here is fundamentally different from the one in Ref.~\cite{wilde2018entanglement}, which is proposed to measure entanglement cost in quantum channel simulation. We leave the connection of the two resource theories to future works. Focusing on our scenario, we can similarly define entanglement nongenerating, nonactivating, and commuting operations and study the corresponding resource theories. Here, we focus on separable channels $\E_{sep}$ that map a separable state $\sigma_{A'C'}$ to a separable state $\sigma_{AC}$, $\E_{sep}(\sigma_{A'C'}) = \sigma_{AC}$. Consider resource free operations as superchannels that map separable operations to separable operations, the channel relative entropy of entanglement can be defined by,
\begin{equation}
	E_{rel}(\N_{AC}) = \min_{\E_{sep}} S_{AB}(\N_{AC}\|\E_{sep}).
\end{equation}
The resource theory based on separable operations measures the entanglement generation ability.
Especially, we consider isometry channels $\V_{AC}(\rho_{A'C'})=V\rho_{A'C'}V^\dag$, with $V^\dag V = I_{A'C'}$. 
The entanglement of isometry channels is lower bounded by the entropy of its all possible subchannels,
\begin{equation}
	E_{rel}(\V_{AC}) \ge \max_{\V \in \{\V_{A}(\rho_{A'}|\rho_{C'}),\V_{C}(\rho_{C'}|\rho_{A'})\} }S_{AB}(\V).
\end{equation}
Here, similar to subsystems of states, we define subchannels from $A'$ to $A$ via $\N_{A}(\rho_{A'}|\rho_{C'}) = \tr_{C}[\N_{AC}(\rho_{A'}\otimes\rho_{C'})]$, and subchannels from $C'$ to $C$ via $\N_{C}(\rho_{C'}|\rho_{A'})=\tr_{A}[\N_{AC}(\rho_{A'}\otimes\rho_{C'})]$. The subchannels $\N_{A}(\rho_{A'}|\rho_{C'})$ and $\N_{C}(\rho_{C'}|\rho_{A'})$ are defined conditioned on the input of the traced out system.
When $A'C'$ are null systems, the equal sign is achieved and it reduce to the entanglement of pure states \cite{PhysRevA.57.1619}. While, the equal sign may not be achieved for general isometry channels, indicating other potential definitions of the entropy of channels.

\emph{Discussion.---}
In this letter, we discuss entropic quantities of channels under the operational task of hypothesis testing. For future works, it is interesting to study the role of entanglement in channel hypothesis testing as the one in channel discrimination \cite{PhysRevLett.102.250501}. 
In our definition, we only consider independent and identical inputs for different uses of the channel. As multipartite entanglement is useful for quantum information processing \cite{PhysRevLett.96.010401,giovannetti2011advances}, it is also interesting to study the most general case where the inputs are jointly entangled. In this work, we also define general resource theories of channels via the relative entropy, and discuss the coherence and entanglement of channels. Completing resource theories of channels is of great importance and is left for future works.

\emph{Acknowledgement.} 
We acknowledge helpful discussions with Xiongfeng Ma, Mark Wilde, Pei Zeng, and Qi Zhao. 
This work was supported by BP plc and by the EPSRC National Quantum Technology Hub in Networked Quantum Information Technology (EP/M013243/1).

\emph{Note added.---} 
Recently, the entropy of channels is independently proposed in Ref.~\cite{gour2018entropy}. In that work, the authors proposed several entropy measures, studied their properties, and investigated the operational meaning in quantum channel merging.

\bibliographystyle{apsrev4-1}
\bibliography{Channeltheory}

\clearpage
\appendix
\widetext



\section{Relative entropies of quantum channels and hypothesis testing}

In this section, we introduce the operational task of hypothesis testing.
Hypothesis testing is a fundamental task for identifying whether the ideal model or initial hypothesis can describe the observed data from the practical device. Classically, the probability that the hypothesis is true is related to the Kullback-Leibler divergence \cite{kullback1951} between the probability distributions of the practical model and the initial hypothesis \cite{press2007numerical}. Specifically, consider a device that generates a random variable $X$. 
For the initial hypothesis, the probability of observing output $x_i$ is $q_i = Q_{X=x_i}(x_i)$; while in practice, the actual probability of outputting $x_i$ is governed by $p_i = P_{x=x_i}(x_i)$.
Given $N\gg1$ independent samples of $X$, the probability that the initial hypothesis is correct can be approximated by $2^{-ND_{KL}(P_X\|Q_X)}$ \cite{cover2012elements}. Here $D_{KL}(P_X\|Q_X)=\sum_i p_i\log(p_i/q_i)$ is the Kullback-Leibler divergence of distributions $P$ and $Q$. Classically, the Kullback-Leibler divergence is also an important measure for other tasks, such as coding theory and Bayesian inference \cite{csiszar2011information,burnham2003model}, and has diverse applications in statistics, machine learning, and physics. 

In quantum mechanics, the to-be-tested device can output general quantum states. Suppose the initial hypothesis or the theoretical guess is that the output state is $\sigma$; while the actual output state is described by $\rho$. When a measurement is performed on a single copy of state $\rho$, with failure probability less than $\varepsilon$, the probability \cite{Wang12} or the $p$-value that the initial hypothesis is true is lower bounded by 
\begin{equation}
	p = 2^{-D_{H}^\varepsilon(\rho\|\sigma)},
\end{equation}
where $D_{H}^\varepsilon(\rho\|\sigma)$ is the one-shot  hypothesis testing relative entropy with smoothing parameter $\varepsilon$ \cite{buscemi2009quantum},
\begin{equation}
	D_H^\varepsilon(\rho\|\sigma) = -\log_2 \min_{Q:0\le Q\le I, \tr[Q\rho]\ge1-\veps}\tr[Q\sigma].
\end{equation}
Given $N\gg1$ independent samples of the state $\rho$, the probability that the initial hypothesis $\sigma$ can reproduce the same classical outputs is approximately lower bounded by $2^{-NS(\rho\|\sigma)}$ \cite{hiai1991f,hayashi2005asymptotics}, where $S(\rho\|\sigma) = \tr[\rho\log(\rho)-\rho\log(\sigma)]$ is the quantum relative entropy of quantum states \cite{umegaki1962conditional}. 

\begin{figure}[ht]\centering
{\includegraphics[width=6cm]{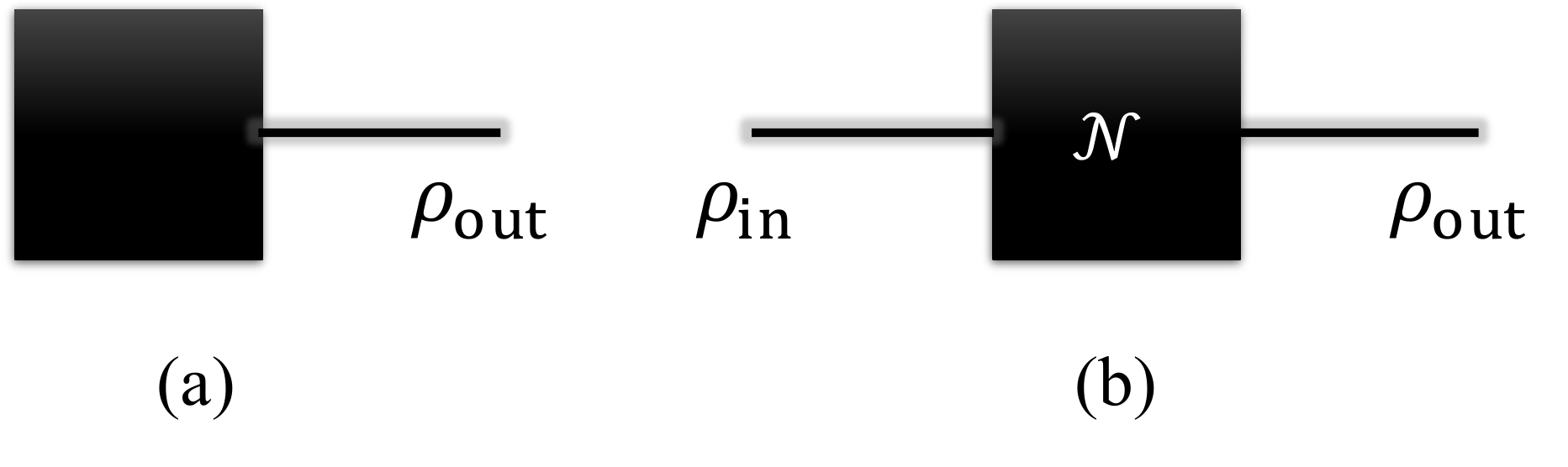}}
 \caption{Quantum states and channels. (a) A quantum state $\rho$ can be regarded as a device that has null input and output state $\rho_{out}=\rho$. (b) A quantum channel is a generalised device that inputs state $\rho_{in}$ and output state $\rho_{out}$. When $\rho_{in}$ has dimension zero or $\rho_{out}$ is a classical state, a quantum channel can be regarded as a state preparation or a demolition measurement, respectively. } \label{Fig:statechannel}
\end{figure}

Briefly speaking, hypothesis testing asks how well the behaviour of a practical device is modelled by the hypothesis. Both a classical variable and a quantum state can be regarded as a device that has null input but only output. While, the most general device in quantum mechanics can have both inputs and outputs. 
As shown in Fig.~\ref{Fig:statechannel},
such a device is called a quantum channel and it also contains the special case of a state preparation device. Focusing on a general quantum channel $\N$, suppose the initial hypothesis assumes that it is described by $\M$. Then hypothesis testing of quantum channels is to investigate how well can the ideal model $\M$ describe the behaviour of the actual channel $\N$ if we use it several times.

\begin{figure}[h]\centering
{\includegraphics[width=8.5cm]{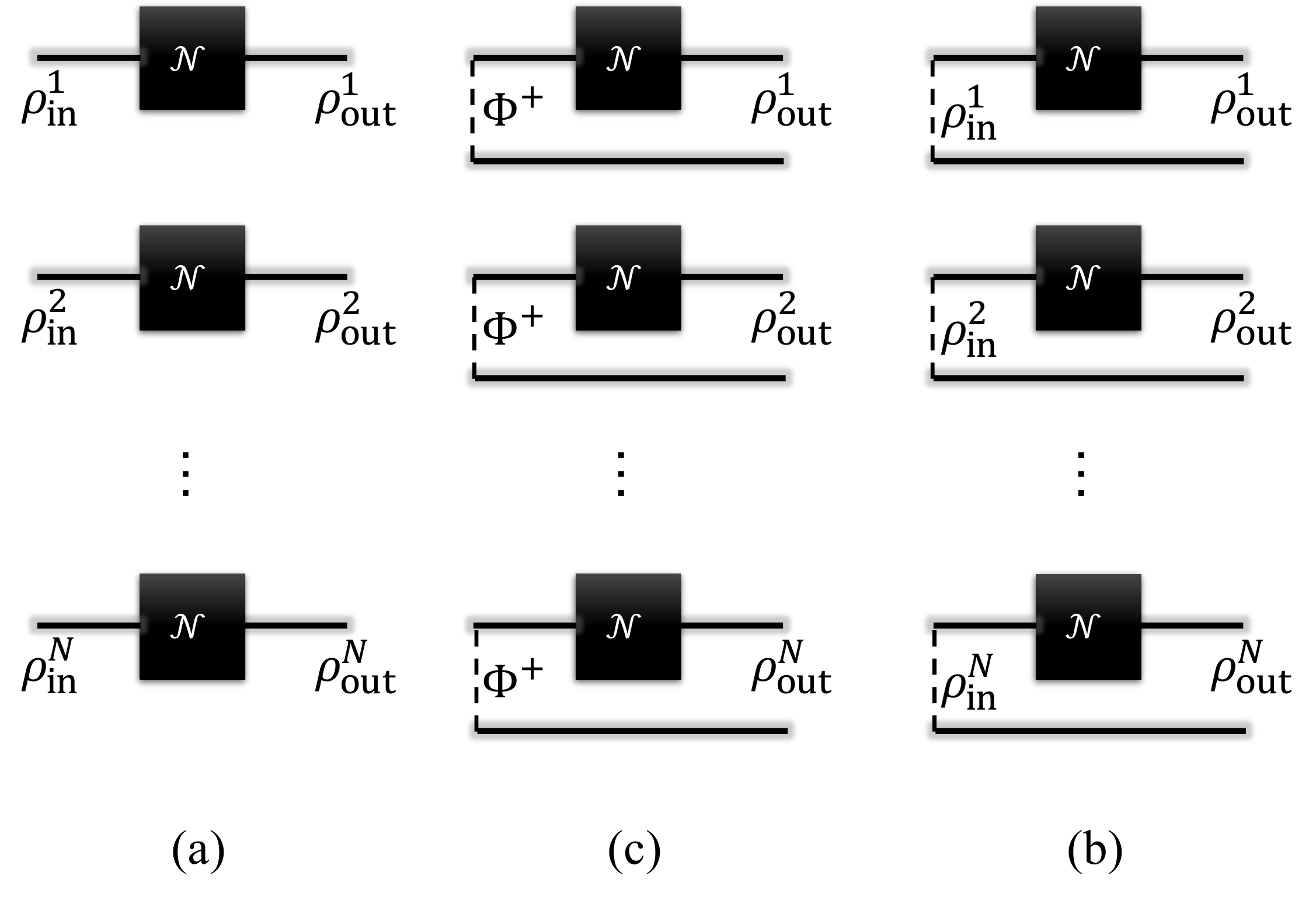}}
 \caption{Hypothesis testing of two channels. (a) For each use of the quantum channel, no extra ancilla is allowed. (b) One party of the maximally entangled state is input to the channel. (c) One party of a joint state is input to the channel.  } \label{Fig:hypothesisSM}
\end{figure}

Focusing on quantum channels $\N$ and $\M$ that both map system $A'$ with dimension $d_{A'}$ to system $A$ with dimension $d_{A}$. Suppose the initial hypothesis is that the quantum channel is $\M$, albeit it is actually described by $\N$. Given $N$ uses of the channel $\N$, we test the correctness of the initial hypothesis. In practice, there are different ways of using the channel by inputting different types of quantum states and performing different measurements of the output states. When the input states of different usages of the channels are independent, we consider three cases that input different types of states as shown in Fig.~\ref{Fig:hypothesisSM}. 

In the first case in Fig.~\ref{Fig:hypothesisSM}(a), we consider that only system $A'$ is input without the help of any ancilla. Although the input state may be entangled with other systems, only system $A$ is available in the output. Denote the input as $\rho_{in}^i$ for the $i^{\textrm{th}}$ usage of the channel, then the output state after $N$ uses of the channel $\N$ is $\N^{\otimes N}(\otimes_i\rho_{in}^i)$. When the hypothesis is true, the output state is $\M^{\otimes N}(\otimes_i\rho_{in}^i)$. Hypothesis testing of channels can thus be reduced to quantum hypothesis testing of the output states with initial hypothesis $\M^{\otimes N}(\otimes_i\rho_{in}^i)$ and actual state $\N^{\otimes N}(\otimes_i\rho_{in}^i)$. The probability that  $\M^{\otimes N}(\otimes_i\rho_{in}^i)$ is true is  $2^{-
	D_{H}^\varepsilon\left(\N^{\otimes N}(\otimes_i\rho_{in}^i)\|\M^{\otimes N}(\otimes_i\rho_{in}^i)\right)}$. As we can choose any input state, the $p$-value that the initial hypothesis $\M$ is true is thus given by minimising overall all possible input states, 
\begin{equation}
	p_{A} = \min_{\rho_{in}} 2^{-
	D_{H}^\varepsilon\left(\N^{\otimes N}(\otimes_i\rho_{in}^i)\|\M^{\otimes N}(\otimes_i\rho_{in}^i)\right)}.
\end{equation}
For the other two cases, we also consider entangled inputs for each use of the channel. In Fig.~\ref{Fig:hypothesisSM}(b), we consider that one party of the maximally entangled state $\Phi^+ = 1/\sqrt{d_{A'}}\sum_i\ket{ii}_{A'B}$ is input to the channel. Although the other ancillary party $B$ is not influenced by the channel $\N$, it is still available in the output state and its entanglement with system $A$ may enhance the hypothesis testing. In Fig.~\ref{Fig:hypothesisSM}(c), we consider that the inputs are all possible states of system $A'$ and any other ancillary system $B$. Denote the $p$-values with $\Phi^+$ and general entangled states as $p_{\Phi^+}$ and $p_{AB}$. It is straightforward that 
\begin{equation}
	p_{AB} \le \min \{p_A, p_{\Phi^+}\},
\end{equation} 
while the order between $p_A$ and $p_{\Phi^+}$ is not determined by definition. In the following, we study the one-shot scenario that the channel is used only once and the asymptotic case that the channel is used infinite times. While as $\N^{\otimes N}$ can be also regarded as one channel, such a one-shot scenario is general for multiple usage of channels. With $N\rightarrow\infty$, the one-shot scenario thus reduces to the asymptotic case.


\section{Properties of quantum relative entropies of states}
In this section, we review the basic properties of the one-shot hypothesis relative entropy $D_H^\varepsilon(\rho\|\sigma)$ and the quantum relative entropy  $S(\rho\|\sigma)$ of states. 
The proof of the properties of $S(\rho\|\sigma)$ can be found in Ref.~\cite{nielsen2010quantum}.
The Non-negativity and monotonicity of $D_H^\varepsilon(\rho\|\sigma)$ can be found in \cite{Wang12}. 
Here, we show that $D_H(\rho\|\sigma)=D_H^{\varepsilon=0}(\rho\|\sigma)$ satisfy the joint convexity and additivity properties.

(Non-negativity) $D_H^\varepsilon(\rho\|\sigma)\ge0$ and $S(\rho\|\sigma)\ge0$. The equality sign for  $D_H^\varepsilon(\rho\|\sigma)$ holds when $\rho=\sigma$ and $\varepsilon=0$. The equality sign for  $D_H^\varepsilon(\rho\|\sigma)$ holds when $\rho=\sigma$.

(Monotonicity) The relative entropy cannot increase by applying any quantum channel $\mathcal{E}$, 
\begin{equation}
	\begin{aligned}
		D_H^\varepsilon(\rho\|\sigma) &\ge D_H^\varepsilon(\mathcal{E}(\rho)\|\mathcal{E}(\sigma)),\\
		S(\rho\|\sigma) &\ge S(\mathcal{E}(\rho)\|\mathcal{E}(\sigma)).
	\end{aligned}
\end{equation}

(Joint convexity) For a set of states $\{\rho_i\}$ and $\{\sigma_i\}$, the relative entropy cannot increase via probabilistically mixing state,
\begin{equation}
	\begin{aligned}
	D_H\left(\sum_ip_i\rho_i\big\|\sum_ip_i\sigma_i\right)&\le \sum_i p_iD_H(\rho_i\|\sigma_i),\\
		S\left(\sum_ip_i\rho_i\big\|\sum_ip_i\sigma_i\right)&\le \sum_i p_iS(\rho_i\|\sigma_i).
	\end{aligned}
\end{equation}

\begin{proof}
	We only focus on mixing two states and the proof can be generalised straightforwardly. Therefore, we need to prove 
	\begin{equation}
		D_H\left(\bar{\rho}\big\|\bar{\sigma}\right)\le p_0D_H(\rho_0\|\sigma_0)+p_1D_H(\rho_1\|\sigma_1),
	\end{equation}
where $\bar{\rho} = p_0\rho_0+p_1\rho_1$ and $\bar{\sigma} = p_0\sigma_0+p_1\sigma_1$.
The definition of the one-shot hypothesis relative entropy with $0$ smoothing is,
\begin{equation}
	D_H(\rho\|\sigma) = -\log_2 \min_{Q:0\le Q\le I, \tr[Q\rho]=1}\tr[Q\sigma].
\end{equation}
Denote the projector onto the space of $\rho$ as $\Pi_\rho$, then 
\begin{equation}
	D_H(\rho\|\sigma) = -\log_2 \tr[\Pi_\rho\sigma].
\end{equation}
Therefore,
\begin{equation}
	\begin{aligned}
		D_H\left(\bar{\rho}\big\|\bar{\sigma}\right) & = -\log_2 \tr[\Pi_{\bar{\rho}}\bar{\sigma}],\\
		&= -\log_2 (p_0\tr[\Pi_{\bar{\rho}}(\sigma_0)]+p_1\tr[\Pi_{\bar{\rho}}(\sigma_1)]),\\
		&\le -\log_2 (p_0\tr[\Pi_{{\rho_0}}(\sigma_0)]+p_1\tr[\Pi_{{\rho_1}}(\sigma_1)]),\\
	   &\le -p_0\log_2 (\tr[\Pi_{{\rho_0}}(\sigma_0)])-p_1\log_2 (\tr[\Pi_{{\rho_1}}(\sigma_1)]),\\
	   &=p_0D_H(\rho_0\|\sigma_0)+p_1D_H(\rho_1\|\sigma_1)
	\end{aligned}
\end{equation}
Here, the second line is due to $\Pi_{\bar{\rho}=p_0\rho_0+p_1\rho_1}\ge \Pi_{\rho_0}$ and $\Pi_{\bar{\rho}=p_0\rho_0+p_1\rho_1}\ge \Pi_{\rho_1}$. This is because whenever $\braket{\psi|\bar{\rho}|\psi} = 0$, it implies that $\braket{\psi|{\rho_0}|\psi}=\braket{\psi|{\rho_1}|\psi} = 0$. The third line is due to the convexity of the $-\log_2(x)$ function.

\end{proof}

(Additivity) For state $\rho_0$, $\rho_1$, $\sigma_0$, $\sigma_1$, the relative entropy is additive
\begin{equation}
	\begin{aligned}
	D_H\left(\rho_0\otimes\rho_1\big\|\sigma_0\otimes\sigma_1\right)&=D_H\left(\rho_0\big\|\sigma_0\right)+D_H\left(\rho_1\big\|\sigma_1\right),\\
		S\left(\rho_0\otimes\rho_1\big\|\sigma_0\otimes\sigma_1\right)&=S\left(\rho_0\big\|\sigma_0\right)+S\left(\rho_1\big\|\sigma_1\right)	\end{aligned}
\end{equation}

\begin{proof}
Denote the projector onto the space of $\rho$ as $\Pi_\rho$, then $\Pi_{\rho_0\otimes\rho_1} = \Pi_{\rho_0}\otimes \Pi_{\rho_1}$ and
	\begin{equation}
		\begin{aligned}
			D_H\left(\rho_0\otimes\rho_1\big\|\sigma_0\otimes\sigma_1\right) &=-\log_2\tr[(\Pi_{\rho_0}\otimes \Pi_{\rho_1})(\sigma_0\otimes\sigma_1)],\\ 
			&=-\log_2\tr[\Pi_{\rho_0}\sigma_0]-\log_2\tr[\Pi_{\rho_1}\sigma_1],\\
			&=D_H\left(\rho_0\big\|\sigma_0\right)+D_H\left(\rho_1\big\|\sigma_1\right).
		\end{aligned}
	\end{equation}
\end{proof}

\section{Properties of quantum relative entropies of channels}
In this section, we investigate the properties of the quantum relative entropy  of channels. 
We refer to $S(\N\|\M)$ the general definition of the relative entropy of two channels.

\paragraph{Pure state is enough in the maximisation}
In the definitions of the quantum relative entropy of channels,  maximisation over all the input state may be required. Here, we show that we only need to consider pure state input. For example, consider the $A$-one-shot relative entropy 
\begin{equation}
	D_{A}({\N}\|{\M}) = \max_\psi D_H(\N(\psi)\|\M(\psi)).
\end{equation}
Suppose the maximisation is achieved with a mixed state $\rho = \sum_ip_i\ket{\psi_i}\bra{\psi_i}$, then 
	\begin{equation}
	\begin{aligned}
		D_{A}(\N\|\M) &= D_H\left(\N\left(\sum_ip_i\ket{\psi_i}\bra{\psi_i}\right)\|\M\left(\sum_ip_i\ket{\psi_i}\bra{\psi_i}\right)\right),\\
		&\le\sum_i p_iD_H(\N(\psi_i)\|\M(\psi_i)),\\
		&\le \max_{\psi} D_H(\N(\psi)\|\M(\psi)).
	\end{aligned}		
	\end{equation}
Note that the proof only requires the joint convexity of the one-shot hypothesis relative entropy of states, it can thus be naturally extended to the other definitions of the quantum relative entropy of channels.

\paragraph{Properties}
In the following, we study the properties of the six definitions of the quantum relative entropies of channels.
\begin{enumerate}
	\item \emph{Non-negativity} The relative entropy is non-negative, i.e., $S(\N\|\M)\ge 0$. The equality sign hods iff $\N\equiv\M$.

By the definitions of quantum relative entropy of channels $S(\N\|\M)$, it is non-negative. When $\N\equiv \M$, it is obvious that $S(\N\|\M) =0$. On the other hand, for $D_A$, when $D_A(\N\|\M) =0$, it implies that $D_H(\N(\psi)\|\M(\psi))=0,\forall \psi$ and hence $\N(\psi)=\M(\psi),\forall \psi$. The proof is similar for $D_{AB}$, $S_A$, and $S_{AB}$. For $D_{\Phi^+}$ and $S_{\Phi^+}$, $D_{\Phi^+}(\N\|\M) =S_{\Phi^+}(\N\|\M) =0$ implies that the Choi matrices of $\N$ and $\M$ are the same, which also implies that $\N\equiv\M$.

\item  \emph{Weak monotonicity}
The relative entropy is non-increasing by sandwiching it with other channels, 
\begin{equation}
	S(\mathcal{X}\circ\N\circ\mathcal{Y}\|\mathcal{X}\circ\M\circ\mathcal{Y})\le S(\N\|\M),
\end{equation}
where $\mathcal{Y}$ maps system $Y$ to $A'$ and $\mathcal{X}$ maps system $A$ to $X$.
		
For $D_{A}$, $D_{AB}$, $S_A$, and $S_{AB}$, it is easy to show $S(\mathcal{X}\circ\N\circ\mathcal{Y}\|\mathcal{X}\circ\M\circ\mathcal{Y})\le S(\mathcal{X}\circ\N\|\mathcal{X}\circ\M)$ as the they are defined by a maximisation over input quantum states. However, for $D_{\Phi^+}$, $S_{\Phi^+}$, the weak monotonicity is not satisfied. This is because one can simply construct $\mathcal{Y}$ such that it swaps the original input state into another state. Due to the example that shows $D_{A}(\N\|\M)>D_{\Phi^+}(\N\|\M)$ and $S_{A}(\N\|\M)>S_{\Phi^+}(\N\|\M)$ for certain channels $\N$ and $\M$, we can thus increase $D_{\Phi^+}$, $S_{\Phi^+}$ by applying the channel to swap the input state to the one that maximise $D_{A}(\N\|\M)$ or $S_{A}(\N\|\M)$. 
For all the six definitions, $S(\mathcal{X}\circ\N\|\mathcal{X}\circ\M)\le S(\N\|\M)$ is true due to the monotonicity of the quantum relative entropy of quantum states.

\item
\emph{Strong monotonicity}
Considering general superchannel transform
\begin{equation}
	\N'_{C'\rightarrow C}=\Phi(\N_{A'\rightarrow A}) = \V'_{AE\rightarrow C}\circ(\N_{A'\rightarrow A}\otimes\I_E)\circ\U_{C'\rightarrow A'E},
\end{equation}
with ancillary system $E$, and channels  $\V'_{AE\rightarrow C}$ and $\V_{AE\rightarrow C}$.
The relative entropy is non-increasing
	\begin{equation}
		S(\Phi(\N)\|\Phi(\M))\le S(\N\|\M).
	\end{equation}		

The proof for $D_{AB}(\N\|\M)$ and $S_{AB}(\N\|\M)$ are similar, so we take $D_{AB}(\N\|\M)$ as an example.
\begin{equation}
	\begin{aligned}
		&D_{AB}(\Phi(\N)\|\Phi(\M))\\
		=&D_{AB}(\V'_{AE\rightarrow C}\circ(\N_{A'\rightarrow A}\otimes\I_E)\circ\U_{C'\rightarrow A'E}\|\V'_{AE\rightarrow C}\circ(\M_{A'\rightarrow A}\otimes\I_E)\circ\U_{C'\rightarrow A'E}),\\
		=&\max_{\psi_{C'B}} D_H((\V'_{AE\rightarrow C}\circ(\N_{A'\rightarrow A}\otimes\I_E)\circ\U_{C'\rightarrow A'E})\otimes \I_B(\psi_{C'B})\|\\
		&\quad \quad \quad \quad \quad (\V'_{AE\rightarrow C}\circ(\M_{A'\rightarrow A}\otimes\I_E)\circ\U_{C'\rightarrow A'E})\otimes I_B(\psi_{C'B})),\\
		\le&\max_{\psi_{A'EB}} D_H((\V'_{AE\rightarrow C}\circ(\N_{A'\rightarrow A}\otimes\I_E))\otimes \I_B(\psi_{A'EB})\|(\V'_{AE\rightarrow C}\circ(\M_{A'\rightarrow A}\otimes\I_E))\otimes I_B(\psi_{A'EB})),\\
		\le&\max_{\psi_{A'EB}} D_H((\N_{A'\rightarrow A}\otimes\I_E)\otimes \I_B(\psi_{A'EB})\|(\M_{A'\rightarrow A}\otimes\I_E)\otimes I_B(\psi_{A'EB})),\\
		=&D_{AB}(\N\|\M).\\
	\end{aligned}
\end{equation}
Here, the third line follows from by replacing the maximisation over $\U_{C'\rightarrow A'E}\otimes \I_E(\psi_{C'B})$ with a larger set of all possible state $\psi_{A'EB}$, the fourth line follows from the monotonicity of $D_H$.
For the other four definitions $D_A(\N\|\M)$, $S_A(\N\|\M)$, $D_{\Phi^+}(\N\|\M)$, $S_{\Phi^+}(\N\|\M)$, general superchannel transform allows the input state coupled with ancilla, which becomes the cases for $D_{AB}(\N\|\M)$ and $S_{AB}(\N\|\M)$. As the examples in the main text show $D_{AB}(\N\|\M)>D_A(\N\|\M)$,  $D_{AB}(\N\|\M)>D_{\Phi^+}(\N\|\M)$, $S_{AB}(\N\|\M)>S_A(\N\|\M)$, and $S_{AB}(\N\|\M)>S_{\Phi^+}(\N\|\M)$ for certain $\N$ and $\M$, the strong monotonicity is not satisfied for $D_A(\N\|\M)$, $S_A(\N\|\M)$, $D_{\Phi^+}(\N\|\M)$, or $S_{\Phi^+}(\N\|\M)$.

\item 
\emph{Joint convexity} The relative entropy is jointly convex 
\begin{equation}
	S\left(\sum_ip_i\N_i\big\|\sum_ip_i\M_i\right)\le \sum_i p_iS(\N_i\|\M_i)
\end{equation}
Here $\N_i$ and $\M_i$ map system $A'$ to system $A$.
		
We first consider $D_A(\N\|\M)$. 
Suppose the state that achieves the maximisation of $D_A\left(\sum_ip_i\N^i\big\|\sum_ip_i\N_2^i\right)$ is $\psi$, then
	\begin{equation}
		\begin{aligned}
			D_A\left(\sum_ip_i\N^i\big\|\sum_ip_i\N_2^i\right) &= D_H\left(\sum_ip_i\N^i(\psi)\big\|\sum_ip_i\N_2^i(\psi)\right),\\
			&\le \sum_ip_iD_H(\N^i(\psi)\| \M^i(\psi)),\\
			& \le \sum_i p_iD_A(\N^i\|\M^i).
		\end{aligned}
	\end{equation}
	The second line follows from the joint convexity of the one-shot hypothesis relative entropy $D_H$ of quantum states. Similarly, the joint convexity is satisfied for all the other definitions.
	
%
%
%

\item 
\emph{Additivity} Suppose $\N_0$ and $\M_0$ evolves system $A'$ to $A$ and $\N_1$ and $\M_1$ evolves system $B'$ to $B$, then the additivity property requires
 	\begin{equation}
 		\Se{\N_0 \otimes \N_1}{\M_0 \otimes \M_1}	 \ge  \Se{\N_0}{\M_0} + \Se{\N_1}{\M_1}.
 	\end{equation}
For relative entropies $D_{A}$ and $S_A$, 
\begin{equation}
	\begin{aligned}
		\Se{\N_0 \otimes \N_1}{\M_0 \otimes \M_1}&=\max_{\psi_{01}} S((\N_0 \otimes \N_1)\psi_{01}\|(\M_0 \otimes \M_1)\psi_{01}),\\
		&\ge \max_{\psi_{0}\otimes\psi_1} S((\N_0 \otimes \N_1)(\psi_{0}\otimes\psi_1)\|(\M_0 \otimes \M_1)(\psi_{0}\otimes\psi_1)),\\
&=\max_{\psi_0}S(\N_0(\psi_0)\|\M_0(\psi_0))+\max_{\psi_1}S(\N_1(\psi_1)\|\M_1(\psi_1)),\\
&=\Se{\N_0}{\M_0} + \Se{\N_1}{\M_1}.
	\end{aligned}
\end{equation}
With $\N_0 = \I$ and $\M_0 = \I$, the equal sign cannot be true due to the example in the main text. Therefore, $D_{A}$ and $S_A$ are not strictly additive. 

Similarly, we can prove the additivity for $D_{AB}$, $S_{AB}$. Whether they are strictly additive is left as an open problem.

We also show that $D_{\Phi^+}(\N\|\M)$, $S_{\Phi^+}(\N\|\M)$ are strictly additive. For $D_{\Phi^+}(\N\|\M)$,
\begin{equation}
	\begin{aligned}
		D_{\Phi^+}(\N_0\otimes\N_1\|\M_0\otimes\M_1)&=D_H((\N_0\otimes\N_1)\Phi_{01}^+\|(\M_0\otimes\M_1)\Phi_{01}^+),\\
		&=D_H((\N_0\otimes\N_1)(\Phi_{0}^+\otimes \Phi_{1}^+)\|(\M_0\otimes\M_1)(\Phi_{0}^+\otimes \Phi_{1}^+)),\\
		&=D_{\Phi^+}(\N_0\|M_0)+D_{\Phi^+}(\N_1\|M_1).
	\end{aligned}
\end{equation}
The proof is similar for $S_{\Phi^+}(\N\|\M)$.

\item \emph{Stability}
The stability requirement says
 	\begin{equation}
		\Se{\I \otimes \N}{\I \otimes \M}	 =\Se{\N}{\M}.
	\end{equation}
	
By definition, 	$D_{AB}$ and $S_{AB}$ satisfy the stability requirements. Meanwhile, $D_{\Phi^+}(\N\|\M)$, $S_{\Phi^+}(\N\|\M)$ also satisfy stability because they are additive. While, due to the example in the main text, $D_{A}$ and $S_A$ are not additive.
\end{enumerate}

\section{Resource theory of channels}
\subsubsection{General result}
In this section, we discuss the general resource theory of channels and show how to use the channel relative entropy to define a general resource measure. 

A general channel resource consists of the definition of free channels, free operations (superchannels) of channels, and the quantitative measures for the resource.  

\begin{itemize}
	\item Free channels:  a resource free channel is denoted as $\C$ and the set of $\C$ is $S_{\C}=\{\C\}$.
	\item Free operations (superchannels) of channels: a superchannel $\Phi$ is free if it maps a free channel to a free channel,
	\begin{equation}
		\Phi(\C) = \C'\in S.
	\end{equation}
	The set of free operations is denoted by $S_{\Phi}$ and it contains the maximal set of free operations. In practice, other physical constraints may be added and the actual set of free operations is $S^P_{\Phi}\in S_{\Phi}$.
	\item Resource measures: a real-valued function of $Q(\N)$ that satisfy the following properties
	\begin{enumerate}
		\item (Non-negativity) It is non-negative for general channels and vanishes for resource free channels, $Q(\N)\ge 0$ and $Q(\C)=0,\forall \C\subseteq S_\C$.
		\item (Monotonicity) It is non-increasing under resource free operations, 
		\begin{equation}
			Q(\N)\ge Q(\Phi(\N)),\forall \Phi\in S_{\Phi}^P.
		\end{equation}
		\item (Convexity) For a set of channels for channels $\{\N_i\}$, it cannot be increased under mixing, 
		\begin{equation}
			\sum_ip_iQ(\Phi(\N_i))\ge Q\left(\sum_ip_i\N_i\right), 		
		\end{equation}
		where $\sum_i p_i=1$. 
	\end{enumerate}
	
\end{itemize}

Now, we define the resource measure via channel relative entropies,
\begin{equation}
	Q_{rel}(\N) = \min_{\C\in S_\C} D(\N\|\C),
\end{equation}
where $D$ is one of the six definitions in the main text.

It is easy to verify the non-negativity requirement as channel relative entropies are also non-negative. 

The monotonicity requirement is only satisfied for $D_{AB}$ and $S_{AB}$ as the other definitions violate the strong monotonicity property. 

	\begin{equation}
		\begin{aligned}
			Q(\N)&=\min_{\C\in S_\C} D(\N\|\C),\\
			& \ge \min_{\C\in S_\C} D(\Phi(\N)\|\Phi(\C))\\
			& = \min_{\C'=\Phi(\C),\C\in S_\C} D(\Phi(\N)\|\C')\\
			& \ge \min_{\C'\in S_\C} D(\Phi(\N)\|\C')\\
			& =Q(\Phi(\N)).
		\end{aligned}
	\end{equation}
Here, the second line follows from the strong monotonicity property of channel relative entropies, the the fourth line is true by minimising over a larger set of $\C'$.	

The convexity is satisfied for all the six definitions.
\begin{equation}
	\begin{aligned}
		\sum_ip_iQ(\Phi(\N_i)) &= \sum_ip_i\min_{\C\in S_\C} \D(\N_i\|\C),\\
		&= \sum_ip_i \D(\N_i\|\C_i),\\
		&\ge  \D\left(\sum_i p_i\N_i\|\sum_i p_i \C_i\right),\\
		&\ge \min_{\C\in S_\C} \D\left(\sum_i p_i\N_i\|\C\right),\\
		&=Q\left(\sum_ip_i\N_i\right).
	\end{aligned}
\end{equation}
Here, in the second line, we denote $\C_i$ to be the channels that achieves the minimisation for $\min_{\C\in S_\C} \D(\N_i\|\C)$. The third line follows from the joint convexity of channel relative entropies. The fourth line follows by replacing $\sum_i p_i \C_i$ with a minimisation over all the resource free set. It is true when $\sum_i p_i \C_i\in S_\C$, that is, the set of $S_\C$ is convex.

\subsubsection{Coherence of general channels}
Consider channels $\N$ that map system $A'$ to system $A$ with computational bases $I_{A'} = \{\ket{{i}_{A'}}\}$ and $I_A = \{\ket{{i}_A}\}$, respectively.
Denote $\Delta_{A'}(\rho_{A'}) = \sum_{i_{A'}}\bra{i_{A'}}\braket{i_{A'}|\rho_{A'}|i_{A'}}\ket{i_{A'}}$ and $\Delta_A(\rho_A) = \sum_{i_A}\bra{i_A}\braket{i_A|\rho_A|i_A}\ket{i_A}$ to be the completely dephasing channels on system ${A'}$ and $A$, respectively. 

Three different types of resource free states are defined \cite{theurer2018quantifying}. 
A channel $\C_d$ is called detection-incoherent when 
\begin{equation}
	\Delta_A\circ\C_d=\Delta_A \circ\C_d\circ \Delta_{A'}.
\end{equation}
A channel $\C_c$ is called creation-incoherent when 
\begin{equation}
	\C_c\circ\Delta_{A'}=\Delta_A \circ\C_c\circ \Delta_{A'}.
\end{equation}
A channel $\C_{dc}$ is called detection-creation-incoherent when 
\begin{equation}
	\Delta_A\circ\C_{dc}=\C_{dc}\circ \Delta_{A'}.
\end{equation}

The channel relative entropy of coherence can be defined by
\begin{equation}
\begin{aligned}
	C_{rel}^d(\N) &= \min_{\C_d} S_{AB}(\N\|\C_d),\\
	C_{rel}^c(\N) &= \min_{\C_c} S_{AB}(\N\|\C_c),\\
	C_{rel}^{dc}(\N) &= \min_{\C_{dc}} S_{AB}(\N\|\C_{dc}).
\end{aligned}
\end{equation}
When resource free operations are operations that maps resource free channels to resource free channels, we can follows the proof in the last section.

However, our definition is slightly different from the resource framework in Ref.~\cite{theurer2018quantifying}, which define resource free operations by a sequential and/or parallel concatenation with resource free channels. Suppose systems $A$ and $B$ are the same, a superchannel $\Phi^{AA'}$ is free when it can be realised by
\begin{equation}
	\Phi^{AA'}(\N) = \C_2^{AA'}\circ (\N^A\otimes \I^{A'}) \circ\C_1^{AA'},
\end{equation}
where $\C_1^{AA'}$ and $\C_2^{AA'}$ are free channels that map system $AA'$ to $AA'$ and $\I^{A'}$ is the identity channel maps from system $A'$ to $A'$. In our definition, we always focus on channels from a fixed system to another fixed system. However, the superchannel $\Phi^{AA'}(\N)$ also enlarges the input and output systems of $\N$. Define a superchannel $\Phi^{A}(\N)$ that traces out system $A'$,
\begin{equation}
	\Phi^{A}(\N) = \tr_{A'} \circ \Phi^{AA'}(\N),
\end{equation}
then it is easy to verify that $\Phi^{A}(\N)$ maps a resource free state to a resource free state. Then the channel relative entropy of coherence is still valid for superchannels $\Phi^{A}(\N)$. That is, the channel relative entropy of coherence satisfy the monotonicity property under $\Phi^{A}(\N)$.

For general super channels $\Phi^{AA'}(\N)$, the monotonicity is also satisfied. We take $C_{rel}^d(\N)$ as an example,
\begin{equation}
	\begin{aligned}
		C_{rel}^d((\N))&=\min_{\C_d^{A}} S_{AB}\left(\N^A\|\C_d^{A}\right),\\
		&=\min_{\C_d^{A}\otimes\I^{A'}} S_{AB}\left(\N^A\otimes\I^{A'}\|\C_d^{A}\otimes\I^{A'}\right),\\
		&\ge\min_{\C_d^{AA'}} S_{AB}\left(\N^A\otimes\I^{A'}\|\C_d^{AA'}\right),\\
		&\ge \min_{\C_d^{AA'}} S_{AB}\left(\C_2^{AA'}\circ (\N^A\otimes \I^{A'}) \circ\C_1^{AA'}\|\C_2^{AA'}\circ\C_d^{AA'}\circ\C_1^{AA'}\right),\\
		&\ge\min_{\C_d^{AA'}} S_{AB}\left(\C_2^{AA'}\circ (\N^A\otimes \I^{A'}) \circ\C_1^{AA'}\|\C_d^{AA'}\right),\\
		&=C_{rel}^d(\Phi^{A}(\N)).
	\end{aligned}
\end{equation}
The second line follows from the stability of $S_{AB}$. 

\subsubsection{Measurement coherence}

Now, we consider the relative entropy of coherence of qubit quantum measurements. 
A measurement $\M_d$ is called detection-incoherent when 
\begin{equation}
	\M_d(\rho) = \sum_k\tr[M_k\rho]\ket{\phi_k}\bra{\phi_k},
\end{equation}
with $M_k = \sum_{i_A} p_{i_A}^k\ket{i_A}\bra{i_A}$. 
A measurement $\M_c$ is called creation-incoherent when 
\begin{equation}
	\M_c(\rho) = \sum_{i_B}\tr[M_{i_B}\rho]\ket{{i_B}}\bra{{i_B}}.
\end{equation}
A measurement $\M_c$ is called detection-creation-incoherent when 
\begin{equation}
	\M_{dc}(\rho) = \sum_{i_B}\tr[M_{i_B}\rho]\ket{{i_B}}\bra{{i_B}},
\end{equation}
with $M_{i_B} = \sum_{i_A} p_{i_A}^{i_B}\ket{i_A}\bra{i_A}$. Here, the POVM elements also satisfy $M_k\ge 0$, and $\sum_kM_k={1}_A$.

We focus on qubit projective measurement, 
\begin{equation}
	\N(\rho) = \braket{\psi_0|\rho|\psi_0}\ket{\psi_0}\bra{\psi_0}+\braket{\psi_1|\rho|\psi_1}\ket{\psi_1}\bra{\psi_1},
\end{equation}
with normalised qubit basis $\{\ket{\psi_0},\ket{\psi_1}\}$. Then we show that the measurement relative entropies of coherence are 
\begin{equation}
\begin{aligned}
	 C_{rel}^d(\N) &\ge C_{\min}(\ket{\psi_0}),\\
	C_{rel}^c(\N) &= C_{rel}(\ket{\psi_0}),\\
	C_{rel}^{dc}(\N) &= C_{rel}(\ket{\psi_0}).
\end{aligned}
\end{equation}
Here $C_{\min}(\ket{\psi_0})$ and $C_{rel}(\ket{\psi_0})$ are the min-entropy and the relative entropy of coherence of state $\ket{\psi_0}$,
\begin{equation}
	\begin{aligned}
		C_{\min}(\ket{\psi_0}) &= H_{\min}(\psi_0^{diag}),\\
		C_{rel}(\ket{\psi_0}) &= H(\psi_0^{diag}).
	\end{aligned}
\end{equation}
Here $H_{\min}$ and $H$ are the min entropy and relative entropy of states, $\psi_0^{diag}=\Delta({\psi_0})$, and $\Delta$ is the completely dephasing channel into the computational basis. 

Denote $\Delta_{\psi}$ to be the completely dephasinng channel into the measurement basis $\{\ket{\psi_0},\ket{\psi_1}\}$,
	\begin{equation}
		\Delta_{\psi}(\rho) = \braket{\psi_0|\rho|\psi_0}\ket{\psi_0}\bra{\psi_0}+\braket{\psi_1|\rho|\psi_1}\ket{\psi_0}\bra{\psi_1}.
	\end{equation}
We first prove for the relative entropy $C_{rel}^d(\N)$ against detection incoherent measurement. 
\begin{proof}

According to the definition, we have 
	\begin{equation}
\begin{aligned}
	C_{rel}^d(\N) &= \min_{\M_d} S_{AB}(\N\|\M_d),\\
	&\ge \min_{\M_d} S_{AB}(\N\circ\Delta_{\psi}\|
	\M_d\circ\Delta_{\psi}),\\
	&= \min_{\M_d} \max_{\psi\in\{\psi_{A'B'}\}}S_{AB}((\N\circ\Delta_{\psi})\otimes \I(\psi)\|(\M_d\circ\Delta_{\psi})\otimes \I(\psi)),\\
	&= \min_{\M_d} \max_{\psi\in\{\psi_0\otimes\psi_B,\psi_1\otimes\psi_B\}}S_{AB}((\N\circ\Delta_{\psi})\otimes \I(\psi)\|(\M_d\circ\Delta_{\psi})\otimes \I(\psi)),\\
	&= \min_{\M_d} \max_{\psi\in\{\psi_0,\psi_1\}}S_{AB}(\N\circ\Delta_{\psi}(\psi)\|
	\M_d\circ\Delta_{\psi}(\psi)),\\
	&= \min_{\M_d} \max_{\psi\in\{\psi_0,\psi_1\}}-\tr[\psi\log_2(\M_d(\psi))],\\
\end{aligned}
\end{equation}
Here, the second line follows from the monotonicity of channel relative entropy; the fourth line follows from the convexity of relative entropy. 
Note that 
\begin{equation}
	\M_d(\rho) = \tr[M_0\rho]\ket{\phi_0}\bra{\phi_0}+\tr[M_1\rho]\ket{\phi_1}\bra{\phi_1},
\end{equation}
then 
\begin{equation}	
	\begin{aligned}
		-\tr[\psi\log_2(\M_d(\psi))] &= -\log_2(\tr[M_0\psi])|\braket{\psi|\phi_0}|^2-\log_2(\tr[M_1\psi])|\braket{\psi|\phi_1}|^2.
	\end{aligned}
\end{equation}
Therefore, 
\begin{equation}
\begin{aligned}
	C_{rel}^d(\N) &=\min_{\M_d} \max_{\psi\in\{\psi_0,\psi_1\}}\left\{-\log_2(\tr[M_0\psi])|\braket{\psi|\phi_0}|^2-\log_2(\tr[M_1\psi])|\braket{\psi|\phi_1}|^2\right\},\\
	&=\min_{\M_d} \max_{\psi\in\{\psi_0,\psi_1\}}\left\{-\log_2(\tr[M_0\psi^{diag}])|\braket{\psi|\phi_0}|^2-\log_2(\tr[M_1\psi^{diag}])|\braket{\psi|\phi_1}|^2\right\},\\
	&=\min_{M_0} \left\{-\log_2(\tr[M_0\psi_0^{diag}])|\braket{\psi_0|\phi_0}|^2-\log_2(\tr[M_1\psi_0^{diag}])|\braket{\psi_0|\phi_1}|^2\right\},\\
	&\ge -\log_2|\psi_0^{diag}|_\infty.
\end{aligned}
\end{equation}
The second line follows from the fact that $M_0$ and $M_1$ are diagonal in the computational basis; the third line is true because  there always exists $\M_d$ and $\M_d'$ such that $\tr[\psi_0\log_2(\M_d(\psi_0))] =\tr[\psi_1\log_2(\M_d'(\psi_1))]$; the last is because $-\log_2(\tr[M_0\psi_0^{diag}]), -\log_2(\tr[M_1\psi_0^{diag}])\ge -\log_2|\psi_0^{diag}|_\infty$. Here, $|\rho|_\infty = \max\{eig(\rho)\}$ is the infinite norm of $\rho$.

Note that $C_{rel}^d(\N)\le C_{rel}^{dc}(\N)$, therefore we have $C_{rel}^d(\N) \le C_{rel}(\ket{\psi_0})$.
\end{proof}

The proof for $C_{rel}^c(\N)$ and $C_{rel}^{dc}(\N)$ are the same, so we only show it for $C_{rel}^c(\N)$.

\begin{proof}

According to the definition, we have 
	\begin{equation}
\begin{aligned}
	C_{rel}^c(\N) &= \min_{\M_c} S_{AB}(\N\|\M_c),\\
	&\ge \min_{\M_c} S_{AB}(\N\circ\Delta_{\psi}\|
	\M_c\circ\Delta_{\psi}),\\
	&= \min_{\M_c} \max_{\psi\in\{\psi_{A'B'}\}}S_{AB}((\N\circ\Delta_{\psi})\otimes \I(\psi)\|(\M_c\circ\Delta_{\psi})\otimes \I(\psi)),\\
	&= \min_{\M_c} \max_{\psi\in\{\psi_0\otimes\psi_B,\psi_1\otimes\psi_B\}}S_{AB}((\N\circ\Delta_{\psi})\otimes \I(\psi)\|(\M_c\circ\Delta_{\psi})\otimes \I(\psi)),\\
	&= \min_{\M_c} \max_{\psi\in\{\psi_0,\psi_1\}}S_{AB}(\N\circ\Delta_{\psi}(\psi)\|
	\M_c\circ\Delta_{\psi}(\psi)),\\
	&\ge  \min_{\sigma} \max_{\psi\in\{\psi_0,\psi_1\}} S_{AB}(\N\circ\Delta_{\psi}(\psi)\|\sigma),\\
	&\ge   \max_{\psi\in\{\psi_0,\psi_1\}} \min_{\sigma}S_{AB}(\N\circ\Delta_{\psi}(\psi)\|\sigma),\\
	&=  C_{rel}(\psi_0)
\end{aligned}
\end{equation}
Here, the second line follows from the monotonicity of channel relative entropy; the fourth line follows from the convexity of relative entropy; the sixth line follows by replacing the minimisation over $\M_c$ with the minimisation over incoherent state $\sigma = \sum_ip_i\ket{i}\bra{i}$; the last two lines follows because $\min_{\sigma}S_{AB}(\N\circ\Delta_{\psi}(\psi)\|\sigma) = C_{rel}(\psi)$ and $C_{rel}(\psi_0) = C_{rel}(\psi_1)$.

Next, we show that $C_{rel}^c(\N) \le C_{rel}(\psi_0)$. Consider a special  creation incoherent measurement as follows, 
\begin{equation}
\D_c(\rho) =\sigma = p_0\ket{{0}}\bra{{0}}+p_1\ket{{1}}\bra{{1}},
\end{equation}
with coefficients $p_0$ and $p_1$ determined later. Then we have
\begin{equation}
	\begin{aligned}
	C_{rel}^c(\N) &= \min_{\M_c} S_{AB}(\N\|\M_c),\\
	&\le \min_{\D_c} S_{AB}(\N\|\D_c),\\
	&=  S_{AB}(\N\circ\Delta_{\psi}\|
	\D_c\circ\Delta_{\psi}),\\
	&= \max_{\psi\in\{\psi_{A'B'}\}}S_{AB}((\N\circ\Delta_{\psi})\otimes \I(\psi)\|(\D_c\circ\Delta_{\psi})\otimes \I(\psi)),\\
	&= \max_{\psi\in\{\psi_0\otimes\psi_B,\psi_1\otimes\psi_B\}}S_{AB}((\N\circ\Delta_{\psi})\otimes \I(\psi)\|(\D_c\circ\Delta_{\psi})\otimes \I(\psi)),\\
	&=  \max_{\psi\in\{\psi_0,\psi_1\}}S_{AB}(\N\circ\Delta_{\psi}(\psi)\|
	\D_c\circ\Delta_{\psi}(\psi)),\\
	&=  \max_{\psi\in\{\psi_0,\psi_1\}}S_{AB}(\psi\|
	\sigma),\\
	&=  C_{rel}(\psi_0).
\end{aligned}
\end{equation}

Similarly we can prove that $C_{rel}^{dc}(\N) = C_{rel}(\psi_0)$. 

\end{proof}

Note that although the set of detection-creation-incoherent measurements set is strictly smaller than the set of creation-incoherent measurements, the coherence with respect to these two sets are the same. However, for the detection-incoherent set, we only have $C_{rel}^{d}(\N) \le  C_{rel}^{dc}(\N)$ and the equal sign is generally not achieved.

\subsubsection{Entanglement of channels}
Now, we extend the entanglement theory to channels. Focus on channels that map systems $AC$ to system $A'C'$, a channel $\C_{AC}^{sep}$ is called separable when it maps any separable state to a separable state, 
\begin{equation}
	\C_{AC}^{sep}(\sigma_{AC}) = \sigma_{A'C'}
\end{equation}
for separable states $\sigma_{AC}$ and $\sigma_{A'C'}$. Resource free operations can be defined by superchannels that map resource free channels to resource free channels. 
Define the channel relative entropy of entanglement by
\begin{equation}
	E_{rel}(\N_{AB}) = \min_{\C_{ent}} S_{AB}(\N_{AB}\|\C_{AC}^{sep}).
\end{equation}
Then it is easy to see that $C_{rel}(\N)$ is an entanglement measure for channels for both definitions of free operations.
Similar to subsystems of states, we define subchannels from $A'$ to $A$ via 
\begin{equation}
	\begin{aligned}
		\N_{A}(\rho_{A'}|\rho_{C'}) &= \tr_{C}[\N_{AC}(\rho_{A'}\otimes\rho_{C'})],\\
		\N_{C}(\rho_{C'}|\rho_{A'})&=\tr_{A}[\N_{AC}(\rho_{A'}\otimes\rho_{C'})].
	\end{aligned}
\end{equation}
The subchannels $\N_{A}(\rho_{A'}|\rho_{C'})$ and $\N_{C}(\rho_{C'}|\rho_{A'})$ are defined conditioned on the input of the traced out system.

For an isometry channels $\N_{AC}(\rho_{A'C'})$, its entanglement is lower bounded by the entropy of its all possible subchannels,
\begin{equation}
	E_{rel}(\N_{AC}) \ge \max_{\N \in \{\N_{A}(\rho_{A'}|\rho_{C'}),\N_{C}(\rho_{C'}|\rho_{A'})\} }S_{AB}(\N).
\end{equation}

\begin{proof}
	First, we prove $E_{rel}(\N_{AC}) \ge \max_{\N_{A}(\rho_{A'}|\rho_{C'}) }S_{AB}(\N_{A}(\rho_{A'}|\rho_{C'}))$. 
	\begin{equation}
\begin{aligned}
	E_{rel}(\U_{AB}) &= \min_{\C_{ent}} S_{AB}(\U_{AC}\|\C_{AC}^{sep}),\\
	&\ge  \min_{\C_{ent}^{sep}} \max_{\psi_{A'E_A}\otimes \psi_{C'E_C}}S_{AB}((\U_{AC}\otimes \I_{E_AE_C})(\psi_{A'E_A}\otimes \psi_{C'E_C})\|(\C_{AC}^{sep}\otimes \I_{E_AE_C})(\psi_{A'E_A}\otimes \psi_{C'E_C}),\\
	&\ge   \max_{\psi_{A'E_A}\otimes \psi_{C'E_C}}\min_{\C_{ent}^{sep}}S_{AB}((\U_{AC}\otimes \I_{E_AE_C})(\psi_{A'E_A}\otimes \psi_{C'E_C})\|(\C_{AC}^{sep}\otimes \I_{E_AE_C})(\psi_{A'E_A}\otimes \psi_{C'E_C}),\\
	&\ge   \max_{\psi_{A'E_A}\otimes \psi_{C'E_C}}\min_{\sigma_{A'E_A}\otimes \sigma_{C'E_C}} S_{AB}((\U_{AC}\otimes \I_{E_AE_C})(\psi_{A'E_A}\otimes \psi_{C'E_C})\|\sigma_{A'E_A}\otimes \sigma_{C'E_C}),\\
	&=\max_{\rho_C'}\max_{\psi_{A'E_A}}S((\U_{AC}\otimes \I_{E_A})(\psi_{A'E_A}\otimes \rho_C)),\\
	&=\max_{\N_A(\rho_A'|\rho_C')}\max_{\psi_{A'E_A}}S(\N_A(\rho_A'|\rho_C')(\psi_{A'E_A})),\\
	&\ge\max_{\N_A(\rho_A'|\rho_C')}\min_{\psi_{A'E_A}}(S(\N_A(\rho_A'|\rho_C')(\psi_{A'E_A}))-S(\tr_A[\N_A(\rho_A'|\rho_C')(\psi_{A'E_A}])),\\
	&=\max_{\N_A(\rho_A'|\rho_C')} S_{AB}(\N_A(\rho_A'|\rho_C'))
\end{aligned}
\end{equation}
Here, the second line follows by considering a smaller set $\psi_{A'E_A}\otimes \psi_{C'E_C}$ of the maximisation over $\psi_{A'E_AC'E_C}$; The third line follows by exchanging the min max.  
The proof follows for $E_{rel}(\N_{AC}) \ge \max_{\N_{C}(\rho_{C'}|\rho_{A'}) }S_{AB}(\N_{C}(\rho_{C'}|\rho_{A'}))$.

\end{proof}

When $A'C'$ are null systems, the equal sign is achieved and it reduce to the entanglement of pure states \cite{PhysRevA.57.1619}. That is, $E(\ket{\psi}_{AB}) = S(\rho_A) = S(\rho_B)$, $\rho_A = \tr_B[\psi_{AB}]$, and  $\rho_B = \tr_A[\psi_{AB}]$.
However, the equal sign may not be achieved for general isometry channels, indicating other potential definitions of the entropy of channels.

\end{document}